\documentclass{elsart}
\usepackage{graphicx}
\begin{document}
\begin{frontmatter}
\title{High-energy Cosmic Rays\thanksref{label1}}
\thanks[label1]{Work supported in part by the U.S. Department of Energy under
DE-FG02 91ER 40626.}
\author{Thomas K. Gaisser \& Todor Stanev}
\address{Bartol Research Institute, University of Delaware\\
Newark, DE 19716}
\maketitle
\begin{abstract}
After a brief review of galactic cosmic rays in the GeV to TeV 
energy range,
we describe some current problems of interest for particles
of very high energy.  Particularly interesting are two features
of the spectrum, the {\it knee} above $10^{15}$~eV and the
{\it ankle} above $10^{18}$~eV.  An important question is
whether the highest energy particles are
of extra-galactic origin and, if so, at what energy the 
transition occurs.  A theme common to all energy ranges is use of nuclear
abundances as a tool for understanding the origin of the cosmic radiation.
\end{abstract}
\end{frontmatter}
\section{Introduction}
The cosmic-ray spectrum falls
steeply, decreasing by approximately a factor
of 50 per decade increase in energy when plotted as $E\phi(E) = dN/d\ln(E)$.
In the lowest energy region the flux is high enough so that
the elemental composition of the primary cosmic-ray nuclei
can be studied by direct observations with detectors lifted
above the atmosphere by balloons or spacecraft.  In the GeV
range, even individual isotopes can be resolved.
In this lower energy range we have the most detailed
information on which to base a model of the origin of
cosmic rays.

\begin{figure}[htb]
\includegraphics[width=14cm]{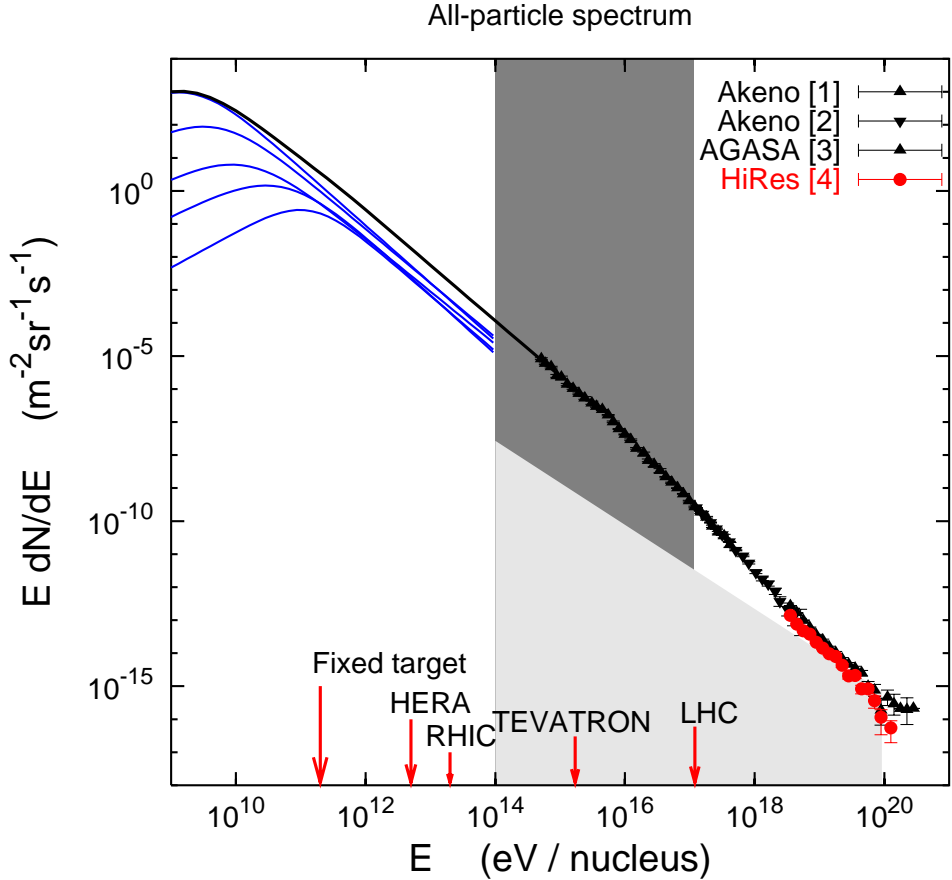}
\caption{Summary of the cosmic-ray spectrum.
The five lines below 100 GeV indicate the individual contributions
of five individual groups of nuclei (protons, helium, CNO, Mg-Si
and Fe).  For clarity only
subsets~\protect\cite{Akeno84,Akeno92,AGASA,HiRes} of the data 
are shown.
}
\label{allpart}
\end{figure}

Above $10^{14}$~eV, where the flux falls below several particles
per square meter per day, direct measurements of the primaries
are no longer practical.  On the other hand, this energy is
large enough so that secondary cascades penetrate
with a footprint large enough to be measured by an array
of detectors on the ground.  Such an extensive air shower (EAS)
array typically has dimensions of a fraction of a square kilometer
or more and can be operated for years rather than days or weeks.
An EAS experiment in effect uses the atmosphere as a
calorimeter, so the showers are classified by total energy
per particle rather than by energy per nucleon as at low energy.
Because of the sparse quality of the information, however, 
the energy per shower is only determined with relatively large
uncertainty, and the
best one can do is to determine the relative contributions
of groups of elements.  The situation is exacerbated by large  
shower-to-shower fluctuations coupled with less than full knowledge
of the properties of the hadronic interactions that govern shower
development. 
The latter problem becomes more severe in the highest energy range,
which is beyond the reach of current accelerators.

Figure \ref{allpart} is a schematic plot of the cosmic-ray spectrum
as a function of total energy per nucleus.  The plot is divided 
into three regions by the shading.  The relatively
detailed measurements of spectra of individual elements (or
groups of elements) below $10^{14}~eV$ supports
a fairly well-developed model of the origin of these particles.
In particular, the observed energy density and propagation time
of the cosmic-rays together determine the power required by the sources.
This connection suggests an approach that may be
helpful in understanding the origin
of the higher energy particles.  The second region (dark shading) includes 
the {\it knee} where there is a steepening and perhaps some structure
in the spectrum that needs explanation.  The 
highest energy region includes the {\it ankle}, 
a flattening of the spectrum above $10^{18}$~eV that may be related to
a transition from particles of galactic origin to those
accelerated in extra-galactic sources.  The question of whether there
is a transition to cosmic rays of extra-galactic origin, and if so
where it occurs, is of considerable interest.
Finally, the question of
whether the spectrum extends beyond $10^{20}$~eV is currently
the foremost problem in high-energy particle astrophysics
because of the difficulty of accounting for the absence of
energy loss by protons in the microwave background radiation
if the spectrum is not suppressed above this energy.  

In what follows we discuss the three energy regions in order.
A common theme is use of the primary composition as a clue to
the nature of the sources of the particles.

\section{Galactic cosmic rays $E\,<\,10^{14}$~eV}

\begin{figure}[htb]
\includegraphics[width=14cm]{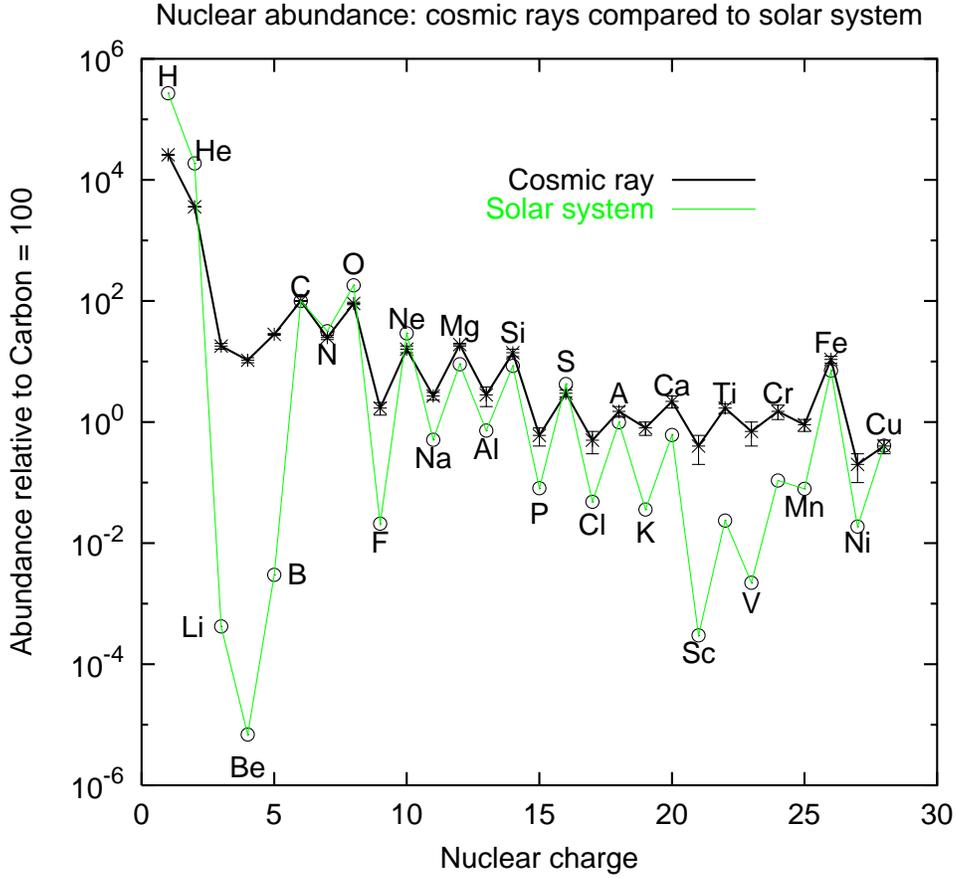}
\caption{Comparison of Solar
system~\protect\cite{solsys} and cosmic-ray elemental 
abundances.  Nuclear abundances are from~\protect\cite{CRabundances}; 
protons and helium are from~\protect\cite{AMSprotons,BESSpHe}.
}
\label{abundances}
\end{figure}

A classic problem in nuclear astrophysics is the
determination of the cosmic-ray source 
abundances~\cite{ShapSilb,Simpson}.
This requires a model of the galaxy, including the spatial 
distribution of cosmic-ray sources; density, composition and 
ionization state of the interstellar medium; and the 
strength and topology of the magnetic field.  Important
recent investigations of cosmic-ray propagation from two
different viewpoints may be found in Refs.~\cite{MoskStrong}
and~\cite{Jones}.  Here we give only the main points in
a simplified form.  The starting
point is a determination of the abundances of different
nuclei near the 
solar system (but outside the heliosphere).  Figure~\ref{abundances}
shows the elemental abundances in the cosmic radiation.
The overabundance by several orders of magnitude
of secondary elements such as lithium, beryllium and boron
relative to their general abundance in solar system
material is a consequence of spallation of the more
abundant primary nuclei, especially carbon and oxygen.
A similar situation occurs for the secondary nuclei
just below iron.  

Given a knowledge of the spallation
cross sections, one has to solve a set of coupled propagation
equations for the diffusion of the cosmic-rays in the turbulent,
magnetized interstellar medium.  A simplified version of the
relevant set of equations is
\begin{equation}
{N_i(E)\over \tau_{esc}(E)}\,=\,Q_i(E)
-\left( \beta c n_H \sigma_i + {1\over \gamma\tau_i}\right)N_i(E)
+\beta c n_H\sum_{k\ge i}\sigma_{k\rightarrow i}\,N_k(E).
\label{propagation}
\end{equation}
Here $N_i(E)$ is the spatial density of cosmic-ray nuclei
of mass $i$, and $n_H$ is the number density of target nuclei (mostly hydrogen) 
in the interstellar medium, $Q_i(E)$ is the number of primary nuclei
of type $i$ accelerated per cm$^3$ per second, 
and $\sigma_i$ and $\sigma_{k\rightarrow i}$
are respectively the total and partial cross sections for interactions
of cosmic-ray nuclei with the gas in the interstellar medium.  The
second term on the r.h.s. of Eq.~\ref{propagation} represents losses due
to interactions with cross section $\sigma_i$
and decay for unstable nuclei with lifetime $\tau_i$.

  Essentially, in the simplest diffusion model, the 
secondary to primary ratios determine the product
of density of the interstellar medium times the characteristic
propagation time of cosmic-rays before they escape from 
the Galaxy ($\tau_{esc}$).  This can be seen by
solving Eq.~\ref{propagation} for a secondary nucleus $S$ while
neglecting its losses during propagation and assuming that 
$Q_S\approx 0$.  Representing the parent primary nuclei by a
single contribution $N_P$ then leads to a simplified equation
\begin{equation}
{m_p\over\sigma_{P\rightarrow S}}\,{N_S\over N_P}\;
=\;\rho\,\beta\,c\,\tau_{esc}\;\approx\;5\;{\rm g/cm}^2.
\label{propagation2}
\end{equation}
The approximate value given in Eq.~\ref{propagation2} requires
accounting for several terms with appropriate spallation cross sections
for its derivation from the data.  
The numerical value in Eq.~\ref{propagation2}
would imply $\tau_{esc}\,\approx\,3\times 10^6$~years
if the average density $\rho$ corresponds to one hydrogen atom per
cubic centimeter as in the disk of the galaxy.

Two details of the data are particularly important.
One is the ratio $R_{10}$ of the  unstable isotope 
$^{10}$Be to stable $^9$Be.  $^{10}$Be is unstable to  
$\beta$-decay with a half-life of $\tau_{10}\;=\;1.5\times 10^6$~years.
The two isotopes of beryllium are produced in
comparable amounts by spallation of heavier nuclei.
If $R_{10}$ as measured is comparable
to the production ratio, then $\tau_{esc}\ll\tau_{10}$; if
$R_{10}\sim 0$ then $\tau_{esc}\gg\tau_{10}$.  The data 
indicate 
\begin{equation}
\tau_{esc}\approx 2\times 10^7\;{\rm years}.
\label{escape}
\end{equation}
Given the mean density of the interstellar medium, the
relatively large value of $\tau_{esc}$
implies that cosmic ray nuclei spend significant time
diffusing in low-density regions, possibly above
the gaseous disk in the galactic halo, before
escaping into inter-galactic space.

The other significant feature is that the ratio of
secondary to primary nuclei decreases with increasing
energy.  From Eq.~\ref{propagation2} this implies that
$\tau_{esc}$ decreases with energy.
This behavior is attributed to energy-dependent diffusion,
whereby higher energy particles diffuse out of the
galaxy more quickly than those of lower energy. 
It should apply to all cosmic rays, including primary protons,
as well as secondary nuclei if the model is consistent
and complete. 
The observed primary spectrum $\phi(E)$ and the source
spectrum $Q(E)$ are connected by a relation of the form
\begin{equation}
\phi(E)\;=\;Q(E)\times\tau_{esc}(E).
\label{source}
\end{equation}
A simple power-law fit to available data gives
\begin{equation}
\tau_{esc}(E) \;\propto\;E^{-\delta}
\label{delta}
\end{equation}
with $\delta\approx 0.6$.  Since the observed differential
energy spectrum is proportional to $E^{-2.7}$, the inferred
source spectrum would be $\propto\,E^{-2.1}$.  

An inferred spectral index at the source of -2.1 is often
cited as evidence for first order diffusive shock acceleration,
which, in the test-particle approximation predicts a value of
-2.0.  The situation is actually more complicated.  On the one
hand, the relation \ref{delta} cannot hold over a large energy
range without coming in conflict with the observed isotropy
of high-energy cosmic-rays.  Extrapolating Eq.~\ref{delta} to
$10^{15}$~eV, for example, would lead to a value of $\tau_{esc}$
almost as small as the light travel time across the galactic disk,
implying a larger anisotropy than is observed.
On the other hand, a more complete treatment of shock acceleration
that accounts for the non-linear back-reaction of the accelerated
particles on the shock structure~\cite{nonlinear}, 
gives a somewhat concave energy
spectrum rather than a single power law (i.e. steeper spectrum 
at low energy, hardening at high energy).  
Models in which some of the energy-dependence of the
secondary/primary ratio is attributed to reacceleration~\cite{reacceleration}
during propagation can accommodate a somewhat steeper average
source spectrum and correspondingly a larger value of
$\tau_{esc}$ at high energy, more consistent with the observed
small anisotropy.  One possibility is that the observed average
spectrum is a composite of contributions from many individual
sources having a range of individual properties.  In what follows,
for illustration we assume $Q(E)\propto E^{-2.4}$ in which case
$\delta\approx 0.3$.

The next step is to estimate the power required of the sources to
maintain the observed spectrum in equilibrium. 
The total source power is related to the locally observed energy spectrum by 
\begin{equation}
\int\,Q(E)\,{\rm d}E\;=\;
{4\pi\over c}\,\int\,E\,{\phi(E)\over\tau_{esc}(E)}\,{\rm d}E.
\label{power}
\end{equation}
The pre-factor on the r.h.s. of this equation converts the measured
cosmic-ray flux $\phi$ to the corresponding spatial density of cosmic rays.
Fixing $\tau_{esc}$ in the GeV range according to Eq.~\ref{escape}
and evaluating the integral using the measured spectrum and
Eq.~\ref{delta} for $\tau_{esc}$
leads to an estimate of the source power as
$\sim 10^{-26}$~erg~cm$^{-3}$s$^{-1}$.
If the flux observed locally near the solar system is typical of
the galactic disc, then the total power of the cosmic-ray
sources is obtained by multiplying by the volume of the disk,
$\sim 10^{67}$cm$^3$, for an estimate of $10^{41}$~erg/s.
The average power generated in kinetic energy of ejected material
(exclusive of neutrinos) in galactic supernova
explosions is $\sim 10^{42}$~erg/s ($10^{51}$~erg/30 year).

The combination of an acceleration theory which produces
approximately the right spectrum with high efficiency (first order diffusive shock
acceleration) and a source with the right power makes supernova shock
acceleration a favored candidate for the origin of cosmic rays in the galaxy.
From observations of synchrotron radiation, it is possible to see
direct evidence of electrons being accelerated to high energy in individual
supernova remnants~\cite{SNRelectrons}.  What 
has proved more difficult is to find examples of individual
supernova remnants in which acceleration of protons can be verified
by the production of neutral pions~\cite{SNRprotons}.  
The problem is that it is
usually possible to explain the observed electromagnetic radiation
as radiation from energetic electrons, by bremsstrahlung at lower 
energy and by inverse Compton scattering
at higher energy.  The characteristic
kinematic signature of pion production (a peak 
at a photon energy of half the pion mass)
tends to be obscured by superimposed synchrotron and bremsstrahlung 
photons~\cite{Butt}.
Indirect arguments related to the shape of the observed gamma-ray spectrum
above a TeV can in some cases most easily be accounted for if the source is
production of $\pi^0$ by protons rather than radiation by 
electrons~\cite{Voelk}.  A nice review of models of galactic
cosmic ray sources is Ref.~\cite{LOD}.

Observation of neutrinos from SNR (or any other potential
cosmic accelerator) would be conclusive evidence for acceleration
of primary protons because only hadronic processes can produce neutrinos.
The expected fluxes are low, however, so large detectors will be needed
to detect them~\cite{neutrinotelescopes}.

\section{The knee: $10^{14} < E < 10^{17}$ eV}

Diffusive shock acceleration works to the extent that charged particles
gain energy by an amount $\Delta E$ proportional to $E$ each time they cross
from upstream to downstream and back upstream of the shock.  (Here
upstream is the unshocked region and $E$ is 
the energy before adding the increment $\Delta E$.)
  After a time $T$ the maximum energy achieved is
\begin{equation}
E_{max}\;\sim\;Ze\beta_s\times B\times T\,V_s,
\label{Emax}
\end{equation}
where $\beta_s = V_s/c$ refers to the velocity of the shock.
This result is an upper limit in that it assumes a minimal diffusion
length equal to the gyroradius of a particle of charge $Ze$
in the magnetic field $B$ upstream of the shock. 
Using numbers typical of Type II supernovae exploding in the average
interstellar medium gives $E_{max}\sim Z\times 100$~TeV~\cite{Cesarsky}.
  More recent
estimates give a maximum energy larger by as much as order of
magnitude or more for some types of supernovae~\cite{Berezhko}.

The nuclear charge, $Z$, appears in Eq.~\ref{Emax} because acceleration depends
on the interaction of the particles being accelerated with the moving
magnetic fields.  Particles with the same gyroradius behave in the same way.
Thus the appropriate variable to characterize acceleration is
magnetic rigidity, $R\,=\,pc/Ze\,\approx E_{tot}/Ze$, where $p$ is the total momentum of
the particle.  Diffusive 
propagation also depends on magnetic fields and hence on rigidity.
For both acceleration and propagation, therefore, if there is a
feature characterized by a critical rigidity, $R^*$, then the corresponding
critical energy per particle is $E^*\,=\,Z\times R^*$.  

\begin{figure}[htb]
\includegraphics[width=14cm]{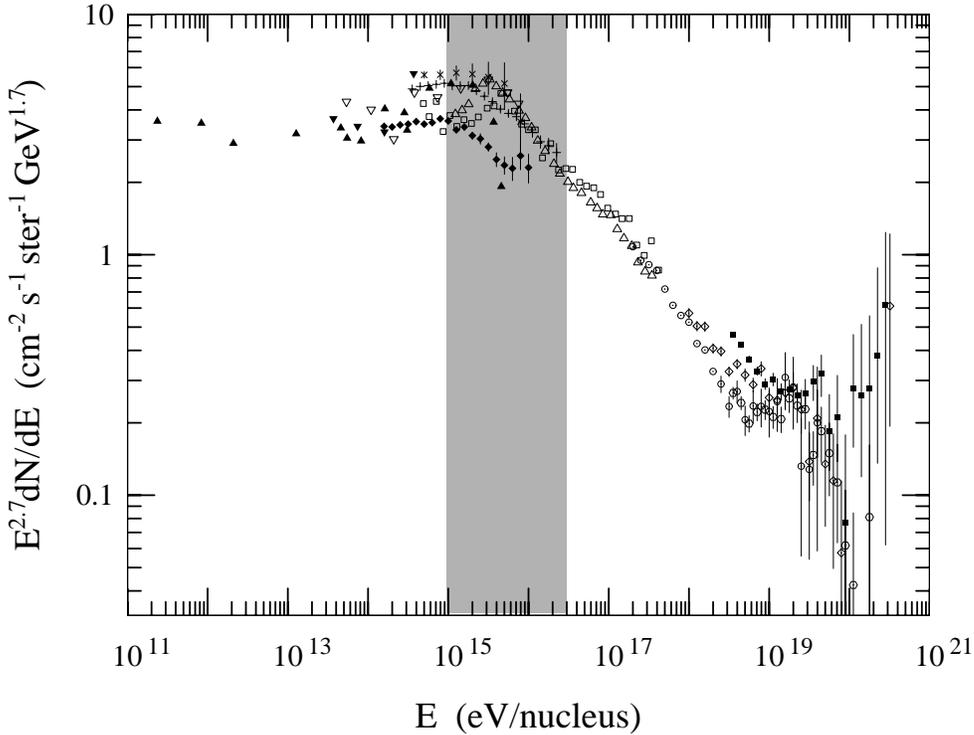}
\caption{High-energy cosmic-ray spectrum.  References to the data
are given in~\protect\cite{rpp}.  The shaded region indicates a
factor a 30 in total energy (see text).
}
\label{Peters}
\end{figure}

The knee of the spectrum is the steepening that occurs above $10^{15}$~eV,
as shown in Fig.~\ref{Peters}, while the ankle is the 
hardening around $3\times10^{18}$~eV.   
One possibility is that the knee
is associated with the upper limit of acceleration by
galactic supernovae, while the ankle is associated
with the onset of an extragalactic population that is less
intense but has a harder spectrum that dominates at sufficiently
high energy.  A generic two-component model was suggested
long ago by Peters~\cite{Peters}, but this explanation
does not fit the data, at least not in its simplest
form.

Consider first the situation if all galactic
sources accelerated particles to the same maximum rigidity, say
$10^{15}$~V.  Then, since the abundant nuclei are in the range
$1\le Z\le 26$, the spectrum would be expected to cutoff
altogether within a factor of about $30$ as a consequence of the
prefactor $Z$ in Eq.~\ref{Emax}.  The protons would cut off
first, followed by helium then CNO, etc.~\cite{Peters}.  The range over 
which the cutoff would occur is indicated by 
the shaded region in Fig.~\ref{Peters}.  Clearly,
this is not at all what happens.  Instead, the spectrum continues
smoothly for another two decades in energy.  Even postulating
a significant contribution from elements heavier than iron 
(up to uranium~\cite{Jorg})
cannot explain the smooth continuation all the way up to the ankle.

One possibility is that most galactic accelerators cut off
around a rigidity of perhaps $10^{15}$~eV, but a few accelerate
particles to much higher energy and account for the region between
the knee and the ankle~\cite{AWW}.  This scenario would be 
a generalization of Peters' model.  Its signature
would be a sequence of composition cycles alternating between
light and heavy dominance as the different components from each
source cut off.
 As emphasized by Axford~\cite{Axford}, however,
the problem with this type of model is that
it requires a fine-tuning of the high-energy spectra so that they rise to 
join smoothly at the knee then steepen to fit the data to
$\sim10^{18}$~eV.  As a consequence, several models have been proposed 
in which the lower-energy accelerators ($E\,<\,10^{15}$~eV) 
inject seed particles into another process that accelerates 
them to higher energy.  In this way the spectrum above the
knee is naturally continuous with the lower energy region.
Groups of supernovae~\cite{Axford} and a termination shock
in the galactic wind~\cite{Jokipii} have been suggested~\cite{Volk}.

The fine tuning problem (i.e. to achieve a smooth spectrum with a sequence of
sources with different maxima) was actually clearly recognized by
Peters in his original statement of this idea~\cite{Peters}.  He correctly
point out, however, that since the cutoff is a function of rigidity
while the events are classified by a quantity close to total energy,
the underlying discontinuities are smoothed out to some extent.
An interesting question to ask in this context is what source power would be 
required to fill in the spectrum from the knee to the ankle.
The answer depends on what is assumed for the spectrum of the sources
and the energy dependence of propagation in this
energy region.  Reasonable assumptions (e.g. $Q(E)\propto E^{-2}$
and $\tau_{esc}\propto E^{-\delta}$ with $\delta\approx 0.3$) lead to an estimate of
$\sim 2\times 10^{39}$~erg/s, less than 10\% of the total power
requirement for all galactic cosmic-rays.  For comparison, the micro-quasar SS433
at 3 kpc distance has a jet power estimated as $10^{39}$~erg/s~\cite{SS433}.

Another possibility is that the steepening of the spectrum at the knee is
a result of a change in properties of diffusion in the interstellar medium
such that above a certain critical rigidity the characteristic
propagation time $\tau_{esc}$ decreases more rapidly with energy.
If the underlying acceleration process were featureless, then
the relative composition as a function of total energy per particle
would change smoothly, with the proton spectrum
steepening first by $0.3$, followed by successively heavier nuclei.
It is interesting that this possibility was also explicitly recognized
by Peters~\cite{Peters}.

A good understanding of the composition would go a long way toward
clarifying what is going on in the knee region and beyond.  
A recent summary of direct measurements of various nuclei shows
no sign of a rigidity-dependent composition change
up to the highest energies 
accessible ($\sim 10^{14}$~eV/nucleus)~\cite{Battiston}.
The change associated with the knee is in the air shower regime.
Because of the indirect nature of EAS measurements, however, the composition is
difficult to determine unambiguously.  The composition has to be determined
from measurements of ratios of different components of air showers at
the ground.  For example, a heavy nucleus like iron generates a
shower with a higher ratio
of muons to electrons than a proton shower of the same energy.  
The best indication at present
comes from the Kascade experiment~\cite{Kascade}, which shows clear evidence
for a ``Peters cycle",
the systematic steepening first of hydrogen, 
then of helium, then CNO and finally the 
iron group.  The transition occurs over an energy range 
from approximately $10^{15}$~eV
to $3\times10^{16}$~eV, as expected, but the experiment runs out of
statistics by $10^{17}$~eV, so the data do not yet discriminate
among the various possibilities for explaining the spectrum between the knee
and the ankle.

\section{The ankle: $E > 10^{17}$ eV}

 The energy range above 10$^{17}$ is where a transition from
 cosmic rays accelerated in the Galaxy to cosmic rays from extra-galactic
sources may be expected.
 The gyroradius of 10$^{19}$ eV protons is of order 10 kpc, exceeding
 the dimensions of the inner Galaxy.  Such particles could 
 not be contained in the Galaxy~\cite{Cocconi56}, not to mention the
difficulty of accelerating particles to such high energy with
known potential galactic accelerators.
 The spectrum of the highest energy cosmic rays indeed 
 shows an `ankle' around $3\times 10^{18}$~eV
 where the power law spectrum flattens to a differential index
 of about $2.7$.  One possibility is that this feature
itself reflects the transition to dominance of extragalactic cosmic rays.
Another possibility is that particles of extragalactic origin
become the dominant population at lower energy and that the ankle
is a feature due to energy loss of protons to electron-positron pair
production on the microwave background during propagation~\cite{Berezinsky}. 

What is needed to decide the question is knowledge of the composition
as a function of energy.
 The idea is that in the range
 where the last galactic accelerator reaches its upper limit the
 composition will start changing in the same way as it does at the `knee'.
 In an energy range of about one and a half decades the composition
 should change from one dominated by protons and light nuclei to 
 heavy nuclei, signaling the upper limit of Galactic sources.
As the extragalactic population of particles begins to dominate, the
composition should change back toward lighter nuclei, assuming the
extragalactic cosmic rays consist primarily of protons.

 \begin{figure}[thb]
\includegraphics[width=14cm]{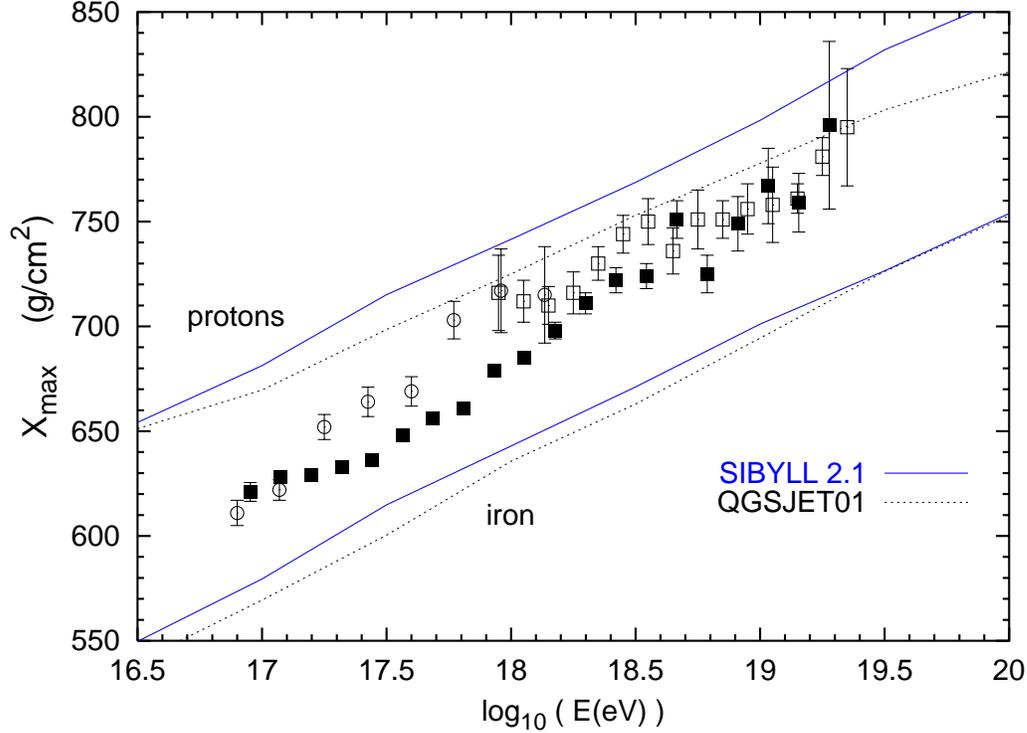}
\caption{Plot of data on mean depth of maximum vs energy.
  Filled squares are data from the stereo Fly's Eye~\protect\cite{FlysEyeXmax}.
 Open symbols show the data of HiRes (squares)
 and HiRes prototype with MIA (circles)~\protect\cite{HiResXmax}.
}
\label{xmax}
\end{figure}

 In this energy range the composition is measured by the energy
 dependence of the position of shower maximum, $X_{max}$.
An air shower consists of a superposition of electromagnetic cascades
initiated by photons from decay of $\pi^0$ particles produced 
by hadronic interactions along the core of the shower as it
passes through the atmosphere.  Most of the energy of the shower
is dissipated by ionization losses of the low-energy electrons and positrons
in these subshowers.  The composite shower reaches a maximum number of
particles (typically $0.7$ particles per GeV of primary energy)
and then decreases as the individual photons fall below the critical
energy for pair production.  Because each nucleus of mass $A$ and total
energy $E_0$ essentially generates $A$ subshowers each of energy $E_0/A$
the depth of maximum depends on $E_0/A$.  Since cascade penetration
increases logarithmically with energy,
\begin{equation}
X_{max}\;=\;\lambda_{ER}\log(E_0/A)\,+\,C,
\label{Xmax}
\end{equation}
where $\lambda_{ER}$ is a parameter (the ``elongation rate") that depends
on the underlying properties of hadronic interactions in the cascade.

 All contemporary interaction models predict an elongation rate between
 50 and 60 g/cm$^2$ per decade of energy for protons.  Showers generated
by iron nuclei would have a similar elongation rate, but would have
$X_{max}$ shallower by $90$ to $100$ g/cm$^2$ at the same total energy
(see Eq.~\ref{Xmax}). 
In general, the depth of maximum should reflect a changing composition
according to Eq.~\ref{Xmax}.  An extreme case would be a composition
changing from pure iron at the end of the galactic component to pure
proton at higher energy if all extragalactic cosmic rays are protons.
In this case $X_{max}$ would increase by $\approx 100$~g/cm$^2$
more than the amount due to the elongation rate alone in the energy
interval in which the transition occurs.
 
Atmospheric fluorescence telescopes such as the Fly's Eye~\cite{FlysE}
directly measure the longitudinal development of large air showers.
After correcting for atmospheric absorption and Cherenkov light
and accounting for energy lost to neutrinos and energetic muons,
the measured shower profile can be fitted and a value of $X_{max}$
determined for each shower~\cite{Song}.
 Figure~\ref{xmax} shows the average 
 depth of shower maximum as measured by three fluorescence experiments.
 The lines show $X_{max}$ predicted by different hadronic
 interaction models~\cite{Sibyll,QGSjet}.
 The model dependence is not high 
 for iron initiated showers, but reaches about 30 g/cm$^2$ for
 very high energy proton showers.

Taken together, the data show a transition from a large fraction
of heavy nuclei around $10^{17}$~eV toward the proton predictions
above $10^{18.5}$~eV.  Whereas the original Fly's Eye stereo 
measurements~\cite{FlysEyeXmax} suggest a mild transition toward protons between
$10^{17.5}$ and $10^{19}$~eV, the more recent HiRes/MIA prototype
data~\cite{HiResXmax} data suggest that the transition begins already
at $10^{17}$~eV and is complete by $10^{18}$~eV.  
The uncertainty about the composition in this region is indicative
of the difficulty of the measurements compounded by uncertainties
in the hadronic interaction models used to interpret the data.
Alan Watson has given a nice review of these problems recently~\cite{Watson}.
He points to progress in solving them as a key to understanding
the origin of the ultra-high energy cosmic radiation.

\begin{figure}[thb]
\includegraphics[width=14cm]{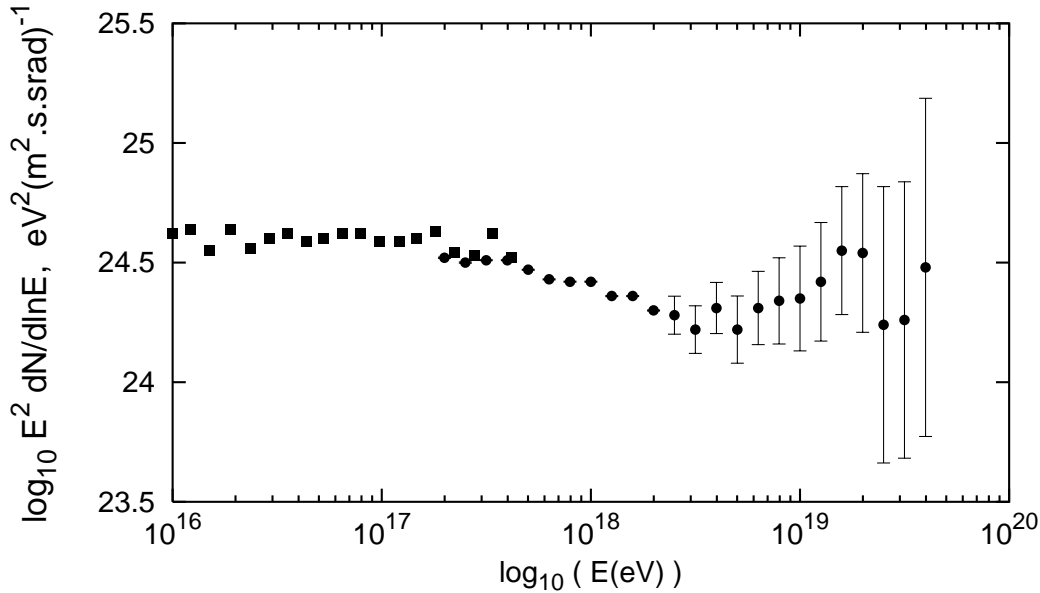}
\caption{ Energy spectrum of the cosmic rays detected by the Fly's Eye
 (filled circles)~\protect\cite{FlysE}. 
 For extension to lower energy the spectrum measured by 
 the Akeno detector is shown with filled squares~\protect\cite{Akeno92}.
}
\label{a_fe}
\end{figure}

Finally we turn to the energy spectrum itself in Fig.~\ref{a_fe}.  
Rather than showing all
data sets as in Fig.~\ref{Peters}, we show only 
the Fly's Eye stereo data at high energy.
 The stereo data consist of air showers detected simultaneously
 by the two Fly's Eye fluorescence detectors. Such events are better
reconstructed and have smaller systematic uncertainties than
showers detected by a single fluorescence telescope.  However,
the exposure for stereo
 observations is significantly smaller and statistical uncertainties
become very large above $10^{19}$~eV.  The advantage of looking at
a single data set is that spectral features become clear.  The
spectrum appears to steepen slightly for $E_0\,>\,10^{17.5}$~eV,
a feature referred to as the ``second knee".  This is followed by
the flattening around $10^{18.5}$~eV which is the ankle. 
Spectra measured by other experiments~\cite{AGASA,HiRes,Yakutsk}
show qualitatively similar features, except for
an apparent excess of events with $E>10^{20}$~eV in 
the AGASA experiment~\cite{AGASA}.
 
If the second knee is associated with the end of the galactic cosmic rays,
then the ankle would most naturally be attributed to pair losses
during propagation in the microwave background, as proposed by
Berezinsky~\cite{Berezinsky}.  In this case, the cosmic rays
down to $\sim10^{17.5}$~eV could be primarily of extragalactic origin.
If instead the ankle is where the
extra-galactic component becomes the dominant population, then most of
the particles could be of galactic origin up to $\sim10^{18.5}$~eV.

The importance of this difference becomes clear when we take the
next step and estimate the power needed to supply the extra-galactic
component.  For a cosmological distribution of sources, the analog
of $\tau_{esc}$ in Eq.~\ref{propagation} is 
the age of the Universe, $\tau_H\approx 10^{10}$~yrs.  The 
required power is then given by Eq.~\ref{power} with
$\tau_{esc}\rightarrow\tau_H$ and with the total
flux $\phi$ replaced by the extragalactic component, $\phi_{EG}$.
The result therefore depends on where the extra-galactic component
is normalized to the observed spectrum and on how it is extrapolated
to low energy below the galactic component.
Assuming a differential 
spectral index of $-2$ and normalizing at $10^{19}$~eV,
the estimated energy density is $2\times 10^{-19}$~erg/cm$^3$.
This leads to an
estimated power requirement of $\sim 10^{37}$erg/Mpc$^3$/s.
Both gamma-ray burst sources (GRB)~\cite{Waxman,Vietri} 
and active galactic nuclei (AGN)~\cite{Berezinsky}
are leading contenders for sources of the ultra-high energy
cosmic rays, in part because their estimated intrinsic
luminosities suggest that they are powerful enough to
satisfy this requirement, given their observed distributions.
Shifting the normalization point down in energy to $10^{18}$~eV
requires a factor 10 more power, which is probably easier to 
satisfy with AGN.  In the GRB model the extragalactic component is
normalized at $10^{19}$~eV~\cite{WB}.

\section{Is there a GZK cutoff?}
This may be the most discussed question in particle 
astrophysics.
It refers to the fact that, for a cosmological distribution
of sources, protons with energy above threshold for
photo-pion production on the microwave background
 are expected to lose energy to this process~\cite{Greisen,ZK}.
Nuclei also lose energy by photo-disintegration.  The result
should be a suppression of the flux for energies above $5\times10^{19}$~eV.
The two experiments with the largest current exposure~\cite{AGASA,HiRes}
show results which give a different indication about whether
the expected suppression is there or not.  This can be seen at
the high end of Fig.~\ref{allpart}.  The main problem is the
extremely low flux~\cite{Olinto}, 
which is of order one particle per square kilometer per
century above $10^{20}$~eV, coupled with the limited exposures
of the detectors.  Answering this question is a major focus of 
the Auger Project~\cite{Auger}.

If the spectrum continues past the GZK energy without suppression,
high-statistics studies will be needed to look for clustering or
anisotropies related to individual sources, which would presumably
be nearby in the absence of a cutoff.  If a suppression is observed,
then accumulation of statistics would be useful to understand the
shape of the spectrum around the GZK feature, which carries
information about the distribution and evolution of sources
on cosmological scales~\cite{Hillas}.  If the spectrum of an extra-galactic
component can be sufficiently well determined also at lower energy
by understanding the sources, then it could be subtracted from
the total observed spectrum revealing the end of the spectrum
of galactic cosmic rays.  Preliminary and speculative versions
of this exercise appear in Ref.~\cite{Berezinsky} and Ref.~\cite{Bergman}.

\end{document}